\documentclass[aps,prl,twocolumn,showpacs,floatfix]{revtex4}
\usepackage[english]{babel}
\usepackage{amssymb}
\usepackage{amssymb}
\usepackage[pdftex]{graphicx}

\newcounter{appsect}
\newcounter{speqno}
\newcounter{aeq}[speqno]

\newcounter{aeqno}
\newcounter{sect}
\begin{document}
\newcommand{\p}{\partial}
\newcommand{\B}{\mathbf}
\newcommand{\h}{\hspace{0.1in}}
\newcommand{\f}{\frac}
\newcommand{\ba}{\begin{array}{llll}}
\newcommand{\ea}{\end{array}}
\newcommand{\be}{\begin{equation}}
\newcommand{\ee}{\end{equation}}
\newcommand{\g}{\gamma}
\newcommand{\dl}{\delta}
\newcommand{\lm}{\lambda}
\newcommand{\s}{\sigma}
\newcommand{\vth}{\vartheta}
\newcommand{\vp}{\varphi}
\newcommand{\G}{\Gamma}
\newcommand{\eps}{\epsilon}
\newcommand{\ra}{\rightarrow}
\newcommand{\Ra}{\Rightarrow}
\newcommand{\Lra}{\Leftrightarrow}
\newcommand{\bel}[1]{\begin{equation}\label{#1}}
\newcommand{\belap}[1]{\begin{appeqa}\label{#1}}
\newcommand{\bsa}[1]{\begin{speqa}\label{#1}}
\newcommand{\esa}{\end{speqa}}
\newcommand{\bsb}[1]{\begin{speqb}\label{#1}}
\newcommand{\esb}{\end{speqb}}
\newcommand{\bsc}[1]{\begin{speqc}\label{#1}}
\newcommand{\esc}{\end{speqc}}
\newcommand{\bsd}[1]{\begin{speqd}\label{#1}}
\newcommand{\esd}{\end{speqd}}
\newcommand{\bse}[1]{\begin{speqe}\label{#1}}
\newcommand{\ese}{\end{speqe}}
\newcommand{\beap}{\begin{appeqa}}
\newcommand{\eeap}{\end{appeqa}}
\newcommand{\bin}[2]{\left(\ba #1 \\ #2 \ea\right)}

\title{Oscillatory instability in super-diffusive 
reaction -- diffusion systems: fractional amplitude and 
phase diffusion equations}
\author{Y. Nec, A.A. Nepomnyashchy}
\affiliation{Department of Mathematics, 
Technion - Israel Institute of Technology, Haifa, Israel}
\author{A.A. Golovin} 
\affiliation{Department of Engineering Sciences and Applied Mathematics, 
 Northwestern University, Evanston, Illinois, USA}
\date{\today}

\begin{abstract}
Nonlinear evolution of a reaction--super-diffusion system near a Hopf 
bifurcation is studied. Fractional analogues of complex Ginzburg-Landau 
equation and Kuramoto-Sivashinsky equation are derived, and some of 
their analytical and numerical solutions are studied.
\end{abstract}
\maketitle

It has been recently realized that in many random physical processes 
the conceptions of Gaussian distribution and Fickian diffusion are 
invalid. Many such processes can be described by models of sub- or 
super-diffusion, where the displacement moments of the corresponding 
random walk grow slower or faster than for normal diffusion, 
respectively. A typical example of {\em super-diffusion} is the enhanced 
transport in fluids, predicted for flows with velocity correlation 
functions slowly decaying in space or time \cite{Bohr}. A specific type 
of super-diffusion, the L\'{e}vy flight, has been reported in 
observations of transport in two-dimensional rotating flows and in a 
freely decaying two-dimensional turbulent flow \cite{Solomon}. Other 
examples of super-diffusive transport include wave turbulence, non-local 
transport in plasma, transport in porous media, surfactant diffusion 
along polymer chains, cosmic rays propagation and motion of animals 
\cite{Balk}. A widely used description of super-diffusive transport 
relies on the continuous time random walk model with a power law 
asymptotics of the particle jump length distribution, leading in the 
macroscopic limit to a diffusion equation with the Laplacian replaced by 
its {\em fractional power} \cite{Montrol}.

An important problem is the influence of super-diffusion on processes 
with chemical reactions \cite{Chen,Brockmann}. Normal reaction-- 
diffusion systems exhibit different types of instability \cite{Murray}. 
Profound understanding of pattern formation and spatio-temporal chaos in 
these systems was achieved through generic equations valid near the 
instability threshold, such as complex Ginzburg-Landau \cite{Cross} and 
Kuramoto-Sivashinsky equations \cite{Bohr}. The evolution of 
instabilities in reaction--diffusion systems can be accompanied by 
advection of components. For instance stirring, which changes the 
effective diffusion properties of species, is one of the means to 
control dynamical regimes generated by instabilities in reaction-- 
diffusion systems \cite{Zhao}. Thus one can expect that in some 
cases flows can give rise to an enhanced diffusion of reagents. While 
studies of instabilities in systems with sub-diffusion have started (see 
\cite{Langlands} and references therein), super-diffusive 
reaction--diffusion systems are still unexplored, with the exception of
the front propagation phenomenon, which is strongly influenced by
fluctuations \cite{Brockmann}. In this letter, weakly non-linear 
dynamics of a reaction--diffusion system characterized by L\'{e}vy 
flights near a long wave bifurcation point is investigated.

Consider a two-component reaction--diffusion system in the general case 
of distinct anomaly exponents:
\bel{act_inh} \f{\p n_j}{\p t}=d_j {\mathfrak D}_{|x|}^{\g_j} n_j
 + f_j(n_1,n_2), \h j=\{1,2\},\ee
where $n_j,d_j$ and $f_j$ are the species concentrations,
diffusion coefficients and general kinetic functions, correspondingly. 
The fractional operator of order $1<\g< 2$ is defined as 
\cite{Sokolov}
\bel{Riesz} {\mathfrak D}_{|x|}^\g n(x)=-\f{\sec(\pi\g/2)}{2\G(2-\g)}
 \f{\p^2}{\p x^2}\int_{-\infty}^{\infty}
 {\f{n(\zeta)}{|x-\zeta|^{\g-1}}d\zeta}.\ee
The equivalent definition in Fourier space allows for a simple 
generalization of the operator to higher spatial dimensions:
$ {\mathfrak D}_{|\B x|}^\g e^{i\, \B q \cdot \B x}=
 -|\B q|^\g e^{i\,\B q \cdot \B x}$ .
Suppose that there exists a homogeneous steady 
state $\B n_0$ satisfying $ \B f(\B n_0)=\B 0$.
A vanishing trace of the sensitivity matrix, 
$(\nabla\B f)_{jk}=\p f_j/\p n_k$, $j,k\in\{1,2\}$,
leads to Hopf bifurcation at the long wave limit $q=0$. Take 
$\eps\ll 1$ and $0<\mu\sim O(1)$ so that 
$\mbox{tr}\,\nabla\B f|_{\B n_0}=\eps^2\mu$ and invoke a multiple 
scales analysis with
$\B n (x,t)=\B N (\xi,t_0,t_2,\ldots;\eps)$,
 $\xi=\dl(\eps) x$, $t_j=\eps^j t, \, j=0,2,\ldots$ and
\bel{asymp_exp_an} \B N \sim \B n_0 +
 \sum_{j=1}^{\infty} \dl_j(\eps) \B N_j(\xi,t_0,t_2,\ldots). \ee
For normal diffusion ($\g=2$) $\dl=\eps$, $\dl_j=\eps^j$ and a 
sequence of problems at successive orders $\dl_j$ is obtained.
The solution at order $\dl_1$ is of the form
$ \B N_1=A(\xi,t_2,\ldots)e^{i\omega_0 t_0}\B v_1 +\mbox{c.c.}$, 
where $\B v_1$ is an eigenvector of the linearized problem and $\omega_0$ is the Hopf 
bifurcation frequency.
Neglecting the phenomena evolving on time scales longer than $\tau=t_2$,
the equation for the amplitude $A$ ensues as a solvability condition at 
order $\dl_3$. For an anomalous system the scaling property of the 
fractional operator,
${\mathfrak D}_{|x|}^\g y(x)=\dl^\g {\mathfrak D}_{|\xi|}^\g y(\xi/\dl)$,
determines the scale of the slow spatial variable, $\dl$. Namely, in a more 
common case with $\g_1=\g_2=\g$, the scale is $\dl=\eps^{2/\g}$ and $\dl_j=\eps^j$.
The amplitude equation has the form 
of a {\em fractional complex Ginzburg-Landau} (FCGL) equation:
\bel{FCGL}\f{\p A}{\p \tau}=A+(1+\alpha i){\mathfrak D}_{|\xi|}^{\g} A-
 (1+\beta i) A |A|^2 \ee
(in rescaled form). This equation was formerly derived in \cite{Tarasov} 
in the problem of nonlinear oscillators' dynamics with long-range 
interactions.
The parameters $\alpha$ and $\beta$ coincide with those of a normal 
reaction -- diffusion system but the Laplacian is replaced by the 
fractional operator.
If $\g_1\neq\g_2$, the super-diffusion term with the larger index is 
negligible in the long-wave region and $\dl_j=\eps^j$ for $j\leq 3$ 
only. Higher-order powers are fractional and depend on the ratio of 
the anomalous exponents. Then the appropriate scaling 
is $\dl=\eps^{2/\g}$ with $\g=\min\{\g_1,\g_2\}$, and the expressions for 
$\alpha$ and $\beta$ are obtained by taking $d_2=0$ if $\g_1<\g_2$ and 
$d_1=0$ if $\g_1>\g_2$.

The integro-differential equation (\ref{FCGL}) retains the basic 
symmetries of a normal complex Ginzburg-Landau equation (with 
respect to time and space translations and the phase change $A 
\mapsto A \exp(i\vth)$). It is interesting that its solutions in the 
form $A(\xi,\tau)=B(\xi)\, e^{i(q\xi-\omega \tau)}$, 
with $q,\omega\in \mathbb R$, have a symmetry similar to that found by 
Hagan \cite{Hagan}. If a solution of this type is known for a pair 
$(\alpha,\beta)$, the solution for a new pair $(\alpha',\beta')$ 
located on one of the curves 
$(\alpha-\beta)/(1+\alpha\beta)=\mbox{const}$ can be found by
the transformation $B=a B', \, \xi=b \xi'$, where
\bsa{ab_p} a^2 b^\g=\f{1+\alpha'\beta'}{1+\alpha\beta}
 \f{1+\alpha^2}{1+{\alpha'}^2}, \esa
\bsb{bp_p} b^\g=\f{1+\alpha^2}{1+\alpha \alpha'+
 (\alpha-\alpha')\omega},\esb
and the new wavenumber and frequency are $q'=bq,\; 
\omega'=\alpha'-b^\g (1+{\alpha'}^2)(\alpha-\omega)/(1+\alpha^2)$. 

In the special case $\alpha=\beta$ eq. (\ref{FCGL}), 
after the phase shift
$A \mapsto A \exp(-i\beta \tau)$, eq.(\ref{FCGL}), 
like a normal complex Ginzburg-Landau equation \cite{Aranson},
can be written in a variational form, 
\be \f{\p A}{\p \tau}=-(1+i\beta)\f{\dl \Upsilon}{\dl A^*}, \ee
where $\Upsilon=\int_{-\infty}^\infty{U(\xi,\tau)d\xi}$, and
$$ U=-|A|^2+\f{|A|^4}{2}-\f{\sec(\pi\g/2)}{2\G(2-\g)}\left\{ 
 \f{\p A^*}{\p \xi}  \f{\p}{\p \xi}\int_{-\infty}^\infty
 {\f{A(\zeta)d\zeta}{|\xi-\zeta|^{\g-1}}}\right. $$
\bel{U_fun} 
 \left. + \f{1-\g}{2}A \int_{-\infty}^\infty{ \f{\p A^*}{\p\zeta}
 \f{\mbox{sign}(\xi-\zeta)}{|\xi-\zeta|^\g}d\zeta}
 +\mbox{\small{c.c.}}\right\}+c.\ee
The constant $c$ is chosen so that $\Upsilon$ converges. Then
$ {\p \Upsilon}/{\p \tau}=-2(1+\beta^2)^{-1}\int_{-\infty}^{\infty} 
 \left| {\p A}/{\p t} \right|^2 d\xi < 0, $
and the system relaxes to a certain stable "stationary" 
solutions (the original variable $A$ oscillates with the frequency 
$\beta$).

Now consider the traveling wave solutions of (\ref{FCGL}),
\bel{A_Riesz} A_q=\sqrt{1-|q|^\g} e^{i(q\xi-\omega \tau)},\,\,
 \omega=\beta-(\beta-\alpha)|q|^\g. \ee
A small perturbation $a(\xi,\tau)$ about $A_q$ comprises 
longitudinal and transverse waves of the form
\be a=A_{q+k}(\tau) e^{i(q+k_\xi)\xi+ik_\eta \eta}+
 A_{q-k}(\tau) e^{i(q-k_\xi)\xi-ik_\eta \eta}, \ee
with $k_\xi,k_\eta$ being the respective wave numbers.
The solution (\ref{A_Riesz}) is neutrally stable with respect to 
disturbances $k_\xi=k_\eta=0$. Further insight into 
long perturbations reveals that for
$ O(k_\xi/q) \sim O(k_\eta/q)\sim o(1)$ to leading 
order the growth rate of $A_{q\pm k}\sim \exp(\lm\tau)$ satisfies
$$ \vspace{-0.2in} \Re \lm \sim \f{\g}{2} |q|^\g \left[-(1+\alpha\beta)
 \left( (\g-1)\f{k_\xi^2}{q^2}+\f{k_\eta^2}{q^2}\right) + \right. $$
\be \left. \hspace{0.4in} \g(1+\beta^2)
 \f{|q|^\g}{1-|q|^\g}\f{k_\xi^2}{q^2} \right]. \ee
Therefore all solutions (\ref{A_Riesz}) are unstable if 
$1+\alpha\beta<0$, i.e. the Benjamin-Feir criterion for a normal CGLE is 
recovered. However, if $1+\alpha\beta>0$, a $\g$-dependent set of 
unstable wave vectors exists, generalizing the Eckhaus instability 
criterion:
\be |q_m|<|q|<1, \h |q_m|^{-\g}=1+\f{\g}{\g-1}
 \f{1+\beta^2}{1+\alpha\beta}.\ee

No new instability criteria emerge in the opposite limit
$q\ll k_\xi,k_\eta\ll 1 $. In particular, the spatially-homogeneous 
oscillation $A_0=\exp(-i\beta\tau)$ is unstable in the same region 
$1+\alpha\beta<0$ with respect to disturbances whose wave numbers $k$ 
satisfy
\bel{BFc} 0<|k|^\g<-\f{2(1+\alpha\beta)}{(1+\alpha^2)},  \h\h
 1+\alpha\beta<0. \ee
The evolution of perturbations near the domain boundary is expected to be 
described by an analogue of the Kuramoto-Sivashinsky equation \cite{Bohr}. 
Define $1+\alpha\beta=-\eps, \h 0< \eps \ll 1$, 
rewrite (\ref{FCGL}) with 
$\chi=\eps^{1/\g} \xi$ and $\tau_2 = \eps^2 \tau$, take
$A=\exp{(-i\beta\tau_2/\eps^2)}\, r(\chi,\tau_2)\,
 \exp{[i\vp(\chi,\tau_2)]}$,
where
$r=1+\sum_{j=1}^{\infty}\eps^j \,r_j(\chi,\tau_2), \;
 \vp=\sum_{j=1}^{\infty}\eps^j \, \vp_j(\chi,\tau_2)$, 
and expand $\exp{(\pm i \vp)}$ to obtain the phase diffusion equation at 
order $O(\eps^3)$ that, after rescaling, has the following form 
(notations for the rescaled space and time variables are the same):
\bel{FKS} \f{\p \phi}{\p \tau}= -{\mathfrak D}_{|\chi|}^\g \phi
 -({\mathfrak D}_{|\chi|}^\g)^2 \phi +
 \f{1}{2}{\mathfrak D}_{|\chi|}^\g \phi^2-\phi
 {\mathfrak D}_{|\chi|}^\g \phi .\ee 
The operator $({\mathfrak D}_{|\chi|}^\g)^2$ is defined in
Fourier space by
$ ({\mathfrak D}_{|\chi|}^\g)^2 e^{iq\chi}=
 |q|^{2\g} e^{iq\chi} $
and cannot be simply related to the operator
${\mathfrak D}_{|\chi|}^{2\g}$ as the order $2\g$ exceeds
the definition range in (\ref{Riesz}). 
Eq.(\ref{FKS}) is the {\em 
fractional 
Kuramoto-Sivashinsky equation}.

\begin{figure}[ht]
\includegraphics[scale=0.4]{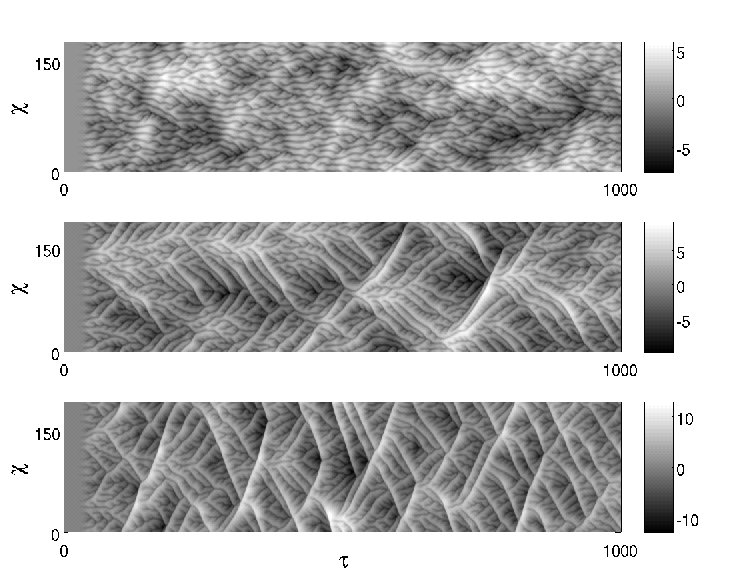}
\caption{Spatio-temporal dynamics of solutions of eq.(\ref{FKS})
for $\gamma=2.0$ (upper), $\gamma=1.7$ (middle), and
$\gamma=1.6$ (lower).}
\label{FKSdynam}
\end{figure}
\begin{figure}[ht]
\includegraphics[scale=0.4]{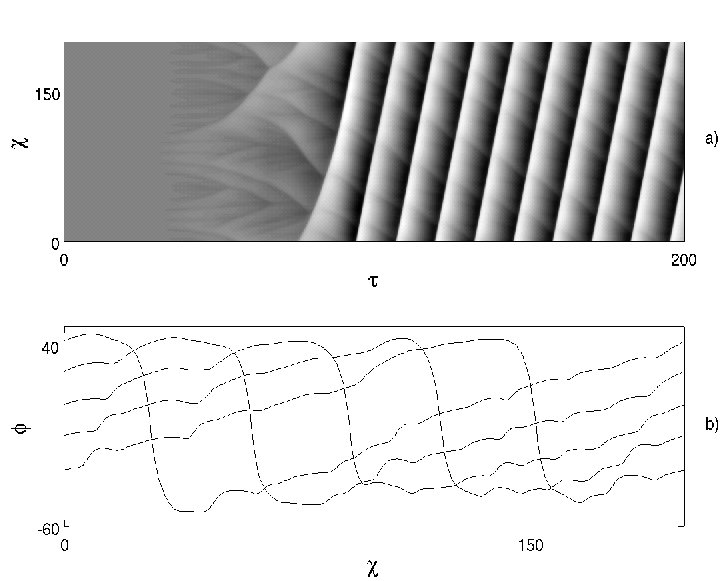}
\caption{(a) Spatio-temporal dynamics of solutions of
eq.(\ref{FKS}) for $\gamma=1.5$.
(b) Solutions of (\ref{FKS}) at successive moments of time.}
\vspace*{-0.2in}
\label{FKSshock}
\end{figure}

\begin{figure}[ht]
\includegraphics[scale=0.4]{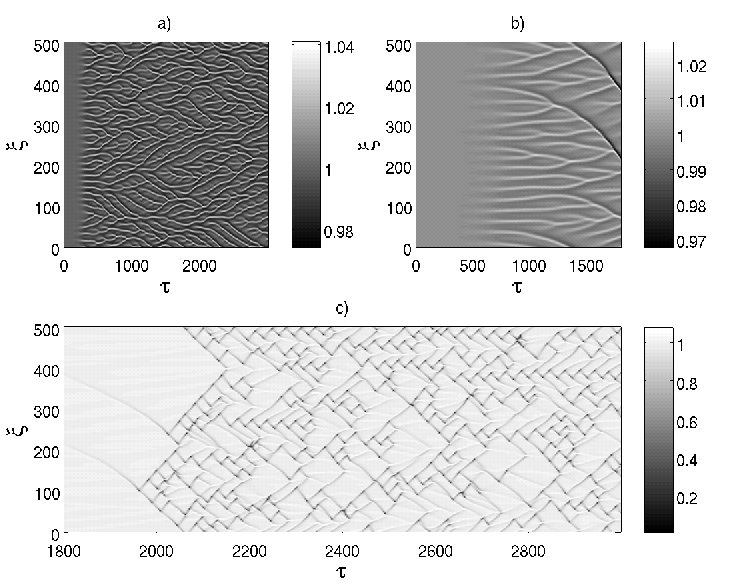}
\caption{Spatio-temporal dynamics of solutions of
eq.(\ref{FCGL}) for (a) $\alpha=-1,\;\beta=1.33,\;\gamma=2.0$;
(b) $\alpha=-1,\;\beta=1.2,\;\gamma=1.6$;
(c) continuation of (b) for the same parameter values.}
\label{FCGLphase}
\end{figure}

Fig.\ref{FKSdynam} shows spatio-temporal dynamics of the numerical 
solutions of eq.(\ref{FKS}) obtained by means of a pseudospectral code, 
using periodic boundary conditions and starting from small-amplitude 
random data. The upper figure shows the dynamics for $\gamma=2$
corresponding to a normal KS equation: this is a well-known
spatio-temporal chaos exhibiting merging and splitting of "cellular"
structures \cite{Bohr}. The middle figure corresponds to $\gamma=1.7$. 
One can see that along with the chaotic dynamics of "cells" 
large-amplitude traveling ``shocks" develop that emit cells still 
displaying chaotic dynamics. With further decrease of $\gamma$, the 
shocks appear more frequently, propagate faster and their amplitude 
grows, see the lower figure corresponding to $\gamma=1.6$. When $\gamma$
decreases below a certain threshold that depends on the domain length, a
single traveling shock is formed in the whole domain. An example of such a
shock is shown in Fig.\ref{FKSshock}. Here the shock is traveling with a 
constant speed (Fig.\ref{FKSshock}a) while its "wings" exhibit 
spatio-temporally chaotic modulations, (Fig.\ref{FKSshock}b). Decreasing 
$\gamma$ results in the increase of the shock amplitude and after 
certain critical $\gamma$ the shock starts accelerating with its 
amplitude growing exponentially. The shock amplitude grows with the size 
of the computational domain. An asymptotic analysis carried out for 
large-amplitude solutions of (\ref{FKS}) shows that the solution is of 
the form $\phi=a(\tau)f(\xi-\zeta(\tau))$, where $f$ is an odd periodic 
function and $a(\tau)$ grows exponentially (despite the problem 
non-linearity) with a certain dependence on the domain size and the 
velocity $d\xi(\tau)/d\tau$ proportional to $a(\tau)$. The numerical 
simulations confirm the asymptotic analysis. 

Next, numerical simulations of the FCGL equation (\ref{FCGL}) in 1D 
have been performed for the phase turbulence regime. 
Fig.\ref{FCGLphase}a shows spatio-temporal dynamics typical of the 
normal CGL equation, starting from the Benjamin-Feir-unstable, 
spatially-homogeneous oscillations: it is well described by the normal 
KS equation (see Fig.\ref{FKSdynam}a). Fig.\ref{FCGLphase}b shows the 
similar dynamics of eq.(\ref{FCGL}) for $\gamma=1.6$. One can see that, 
after some period of phase turbulence, accelerating shocks form that 
trigger the transition to defect turbulence shown in 
Fig.\ref{FCGLphase}c. The formation of the accelerating shocks seen in 
Figs.\ref{FCGLphase}b,c is consistent with the formation of shocks in 
the FKS equation discussed above.
\begin{figure}[ht]
\includegraphics[scale=0.4]{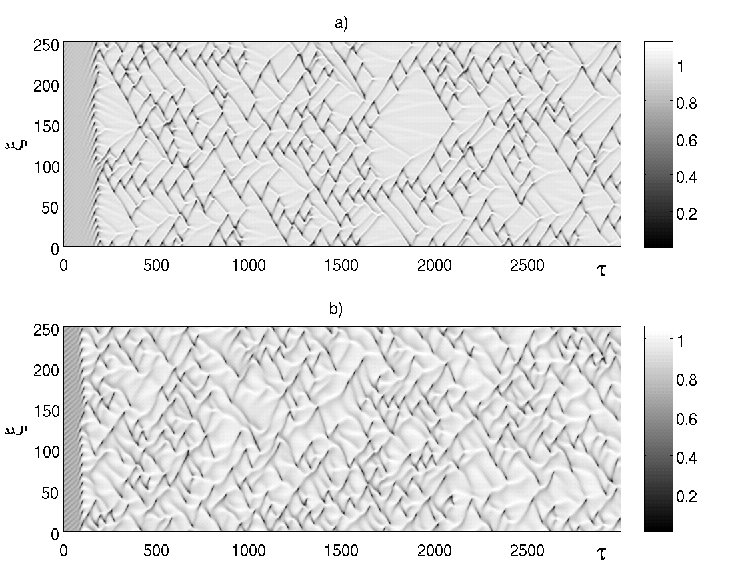}
\caption{Spatio-temporal dynamics of $|A|$ -- solution of
eq.(\ref{FCGL}) for $\alpha=1,\;\beta=-1.3$ and 
(a) $\gamma=2.0$; (b) $\gamma=1.1$.}
\vspace*{-0.2in}
\label{FCGLA}
\end{figure}

Fig.\ref{FCGLA} shows the spatio-temporal dynamics of numerical 
solutions of eq.(\ref{FCGL}) corresponding to the defect turbulence 
regime emerging from the Benjamin-Feir-unstable wave (\ref{A_Riesz}) 
with $q=0.5$ for $\gamma=2.0$ and $\gamma=1.1$ (Figs.\ref{FCGLA} (a) and 
(b), respectively). One can see that in the anomalous case the defect 
turbulence has a stronger phase-turbulence component and does not 
consist of propagating holes.
\begin{figure}[b]
\vspace*{-0.2in}
\includegraphics[scale=0.4]{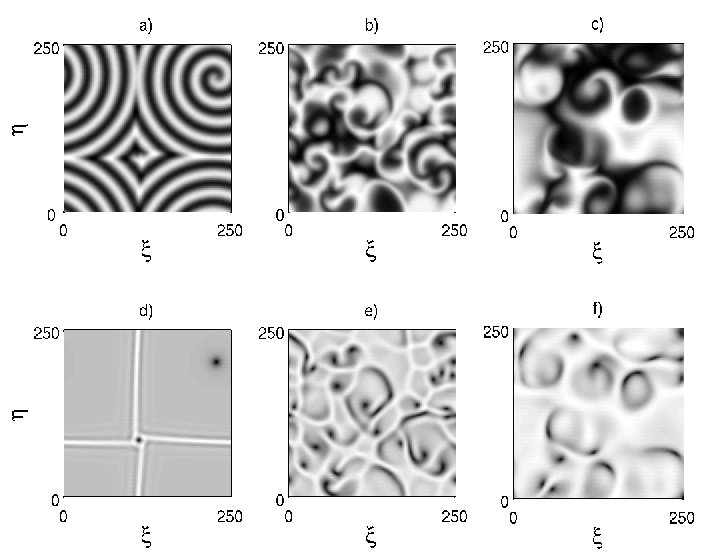}
\caption{Snapshots of solutions of eq.(\ref{FCGL}) for
$\alpha=1.5,\; \beta=-0.6$ and $\gamma=1.9$ (a),(d); $\gamma=1.8$ (b),(e);
$\gamma=1.05$ (c),(f); upper figures -- Re($A$), lower ones - $|A|$.}
\label{FCGLspiral}
\end{figure}

Finally, numerical simulations of FCGL eq.(\ref{FCGL}) in 2D have been 
performed for the parameter values corresponding to the formation of 
spiral waves in the normal CGL equation. Periodic boundary conditions 
and small-amplitude random initial data were used. The results are shown 
in Fig.\ref{FCGLspiral}. One can see that for $\gamma$ close to 2 (see
Figs.\ref{FCGLspiral}a,d) the formation of a spiral wave is still 
observed. With the decrease of $\gamma$ the spiral-wave regime is
replaced by a defect chaos, however, remnants of the spiral waves still 
can be seen (Figs.\ref{FCGLspiral}b,e), with each "spiral" occupying
a small domain with the domain walls partially melted. Further decrease 
of $\gamma$ results in the decrease of the number of defects, the domain 
walls are almost completely melted (Figs.\ref{FCGLspiral}c,f), and the 
local wavenumber created by each defect decreases.
A "phase diagram" of the new dynamical states described above in the 
parameter space will be presented elsewhere.

In conclusion, we have derived fractional Ginzburg-Landau and 
Kuramoto-Sivashinsky equations that describe weakly non-linear dynamics 
of a super-diffusive reaction--diffusion system, characterized by 
L\'{e}vy flights, and studied some of their solutions analytically and 
numerically. We have shown that super-diffusion can lead to a transition from 
phase- to defect turbulence and to destruction of spiral waves. We note 
that investigating the effects of fluctuations on non-linear dynamics of 
instabilities in a super-diffusive reaction--diffusion system would be 
of interest since, as shown in \cite{Brockmann}, the fluctuations can 
have a profound influence on the non-linear behavior in such systems.
However, this topic is beyond the scope of the present paper.

A.A.N. acknowledges the support 
of the ISF grant \#812/06, Minerva Center for Nonlinear Physics of 
Complex Systems and the Technion V.P.R. fund. A.A.G. acknowledges the 
support of the NSF grant \#DMS-0505878.


\vspace*{-0.2in}
\begin{thebibliography}{99}

\bibitem{Bohr} T. Bohr, M.H. Jensen, G. Paladin and A. Vulpiani, {\em 
Dynamical systems approach to turbulence}, 1998, Cambridge Univeristy 
Press.

\bibitem{Solomon} T.H. Solomon, E.R. Weeks and H.L. Swinney, Phys. Rev. 
Lett. {\bf 71}, 3975 (1993); A.E. Hansen, D. Marteau and P. Tabeling, 
Phys. Rev. E {\bf 58}, 7261 (1998).

\bibitem{Balk} A.M. Balk, J. Fluid Mech. {\bf 467}, 163 (2002);
A.V. Chechkin, V. Yu. Gonchar and M. Szydlowski, Phys. Plasmas {\bf 9}, 
78 (2002); D. del Castillo-Negrete, Phys. Plasmas {\bf 13}, 082308 
(2006); M.M. Meerschaert and C. Tadjeran, J. Comp. Appl. Math. 
{\bf 172}, 65 (2004); R. Angelico, A. Ceglie, U. Olsson, G. Palazzo 
and L. Ambrosone, Phys. Rev. E {\bf 74}, 031403 (2006); G. Wilk and 
Z. Wlodarczyk, Nuc. Phys. B - Proc. Suppl. {\bf 75}, 191 (1999);
F.G. Schmitt and L. Seuront, Physica A {\bf 301}, 375 (2001); 
G.M. Viswanathan, V. Afanasyev, S.V. Buldyrev, E.J. Murphy, P.A. Prince 
and H.E. Stanley, Nature {\bf 381}, 413 (1996).

\bibitem{Montrol} E.W. Montroll and G.W. Weiss, J. Math. Phys. {\bf 6},
167 (1965); R. Metzler and J. Klafter, Phys. Rep. {\bf 339},
1 (2000); G.M. Zaslavsky, {\em Hamiltonian chaos and
fractional dynamics}, 2005, Oxford University Press.

\bibitem{Chen} R. Mancinelli, D. Vergni and A. Vulpiani, Physica D
{\bf 185}, 175 (2003); D. del Castillo-Negrete, B.A. Carreras and V.E.
Lynch, Phys. Rev. Lett. {\bf 91}, 018302 (2003).

\bibitem{Brockmann} D. Brockmann and L. Hufnagel, Phys. Rev. Lett. 
{\bf 98}, 178301 (2007).

\bibitem{Murray} J.D. Murray, {\em Mathematical Biology}, 1989,
Springer-Verlag,  New York.

\bibitem{Cross} M.C. Cross and P. Hohenberg, Rev. Mod. Phys. {\bf 65},
851 (1993).

\bibitem{Zhao} B. Zhao and J. Wang, J. Phys. Chem. A {\bf 109}, 3647 
(2005).

\bibitem{Langlands} T.A.M. Langlands, B.I. Henry and S.L. Wearne, J. 
Phys.: Cond. Matt. {\bf 19}, 065115 (2007).

\bibitem{Sokolov} I.M. Sokolov, A.V. Chechkin and J. Klafter, Physica A: 
Stat. Theor. Phys. {\bf 336}, 245 (2004).

\bibitem{Tarasov} V.E. Tarasov, G.M. Zaslavsky, Chaos: {\bf 16}, 023110
(2006).

\bibitem{Hagan} P.S. Hagan, SIAM J. Appl. Math. {\bf 42}, 762 (1982).

\bibitem{Aranson} I.S. Aranson and L. Kramer, Rev. Mod. Phys. {\bf 74},
99 (2002).



\end{thebibliography}
\end{document}